# Analytical Effective Method for Verification of a Satellite Pass over a Region of the Earth Surface

Atanas Marinov Atanassov

*Solar Terrestrial Influences Institute, Bulgarian Academy of Sciences,*
*Stara Zagora Department, P.O. Box 73, 6000 Stara Zagora, Bulgaria*

***Abstract.*** An analytical method is proposed in this work for verification whether an artificial earth satellite during its orbital motion passes over a region of the earth surface. The method is based on undisturbed Keppler's approximation of the orbit and approximation of the region by a circular segment S. In order to define the situational condition, a conic surface is used with apex in the earth centre, cutting out the circular segment. The tangents of the conical surface with Keppler's plane determine the time intervals in which the satellite trace on the earth surface occurs inside the segment S. The transformation of these tangents in the plane of Keppler's orbit and the determination of their crossing points with Keppler's ellipse lies in the basis of the examined method.

## 1. Introduction.

A number of cases exist when, during space experiments, it is necessary to know the time of a satellite pass over a definite region of the earth surface. Thus, for example, in synchronous satellite and ground-based measurements, it is important when the satellite passes over a definite territory where the ground-based station is located. When problems of meteorological character are solved on the basis of satellite information, it is significant when the satellite is going to pass over a definite territory or a meteorological structure (cyclone centre, front). The solution of many other problems, connected with the study of the earth surface from space is connected with the determination of the temporal interval pass over a specific region. This is necessary in some of the cases for experiments planning. In other cases, the analysis is needed to schedule the seances for receiving satellite information. In both cases this is important for the quality of the conducted experiments, and from economical point of view.

The problem for determining a satellite pass over a definite geographic region has a standard solution. It is obtained on the basis of the imitation modelling by selecting a proper geometrical model for region V which determines the situational condition. The discretization of the solution of the artificial earth satellite motion equation and the respective analysis, as concerns the model of the region, allow to determine whether the satellite passes over the region as well as the moments of crossing its borders.

For the equation of the artificial earth satellite motion in geo-equatorial co-ordinate system (GeCS) we have:

(1) $\quad m\dfrac{d^2 \vec{r}}{d t^2} = -\Sigma \vec{f}_k \quad ,$

with initial conditions $\vec{r}_0 = \vec{r}(t_0)$, $\dfrac{d\vec{r}}{dt} = \dfrac{d\vec{r}(t_0)}{dt}$, where $\vec{r}$ is the satellite radius-vector; m - its mass and t- the time. The specific form of (1) reflects the accepted motion model. The solution of (1) can be obtained on the basis of analytical or numerical methods [1,2]. In any case, a discretization of the solution of (1) is obtained:

(2)     $\vec{r}_{t_0}, \vec{r}_{t_1}, \vec{r}_{t_2}, \ldots, \vec{r}_{t_n}, \ldots$

Usually (2) is obtained in GeCS or in orbital co-ordinate system (OCS). It is necessary to transform the solution of (1) into Greenwich co-ordinate system (GrCS):

(3)     $\vec{r}_{(GrKS)} = \alpha_{GrG} \cdot \vec{r}_{(GKS)}$

In (3) $\alpha_{GrG}$ is the transformation matrix [3].

Problems exist in which region V is restricted by a complex outline contour (for example, a state border). There are known methods to present V and to solve the problem for crossing its borders by the sub-satellite trace [4]. Within the terms of different problems, the approximation of region V by a circular spherical segment of the earth surface is completely sufficient and substantiated both physically and of geometrical point of view. The application of such a simplifying situational condition in the discretization of the solution of the artificial earth satellite motion equation requires also considerable computation time.

**2. Formulation of the Problem.**

We shall examine the considered region of the earth surface as a spherical segment S (Fig. 1). It is cut out of the earth surface by a straight circular cone with angle $\psi$ between the axis and the generant and its apex is in the earth centre. The crossing point of the cone axis with the earth surface has Greenwich co-ordinates $(\lambda, \Theta)$. Therefore, the segment can be described by the following parameters – angle $\psi$, earth radius $R_\oplus$ and the Greenwich co-ordinates $\lambda$ and $\Theta$, i.e. $S(\Psi, R_\oplus, \lambda, \Theta)$. Moving along with the earth surface, the cone tangents with the plane of Keppler's orbit at its two sides at moments $t_1$ and $t_2$. (Fig. 2). Between the two moments $t_1$ and $t_2$, the Keppler's plane and the conic surface intercross. This means that part of the Keppler's ellipsis is also restricted within the limits of the conic surface and that it is located over segment S.

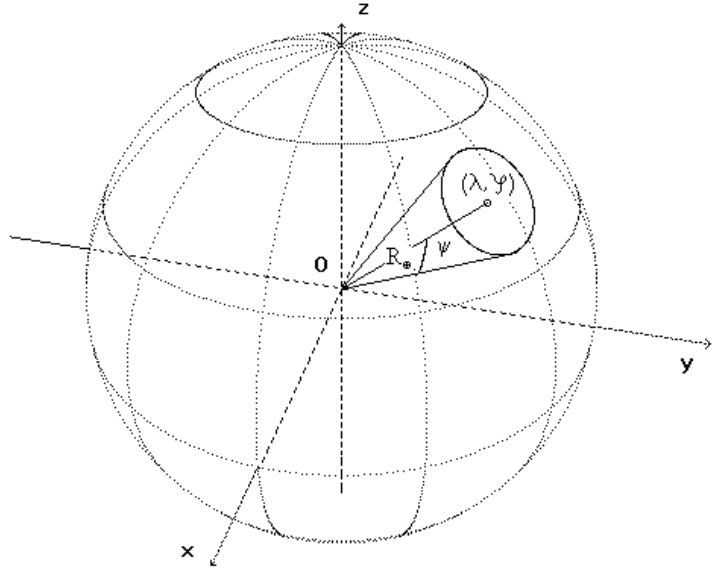

**Figure 1.** A region of the earth surface, presented by a circular segment.

We shall discuss an approach, allowing to obtain moments $\tilde{t}_1$ and $\tilde{t}_2$ when the satellite crosses the cone generants $\vec{\tau}_1$ and $\vec{\tau}_2$ which tangent with the Keppler's orbit.

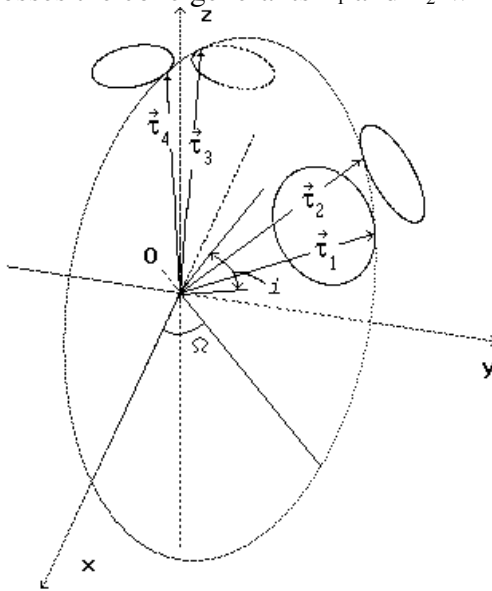

**Figure 2.** The spherical segment crosses Keppler's plane between moments $t_1$ and $t_2$ ($t_3$ and $t_4$, respectively); vectors $\vec{\tau}_1$ и $\vec{\tau}_2$ ($\vec{\tau}_3$ and $\vec{\tau}_4$, respectively) determine the generants, by which the conic surface tangents with Keppler's plane.

The relation between the intervals $(t_1, t_2)$ and $(\tilde{t}_1, \tilde{t}_2)$ on the time axis shows whether the artificial earth satellite passes over segment S (Fig. 3). If the two intervals intercross, then the condition for passing over the examined segment is fulfilled.

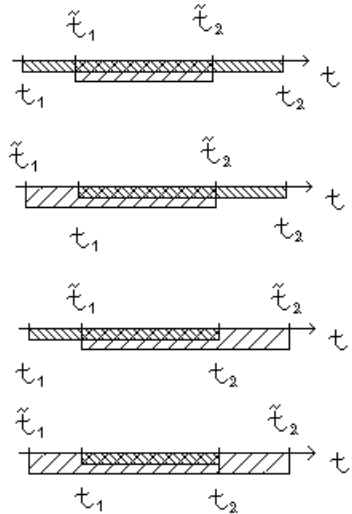

**Figure 3.** Different cross-sections between intervals $(t_1, t_2)$ and $(\tilde{t}_1, \tilde{t}_2)$, in which the situational condition is executed.

### 3. Construction of an algorithm.

Let's assume that segment S forms a tangent with K. For distance $\delta$ from the centre of S to K we can write down [5]:

(4) $\quad \dfrac{\vec{n}}{|\vec{n}|}(\vec{R}_c - \vec{x}) = \delta = \sin\psi \cdot R_\oplus$

or

(4') $\quad \vec{n}^0 \cdot \vec{R}_c = \sin\psi \cdot R_\oplus$

where $\vec{n}^0$ is the null vector of K, $\vec{R}_c$ -is the radius-vector of the segment middle and $R_\oplus = |\vec{R}_c|$ - the Earth radius. The radius-vector of the spherical segment centre $\vec{R}_c$ can be presented in the following way:

(5) $\quad \begin{vmatrix} X_c = R_\oplus \sin\Theta \cdot \cos[\omega_z(t - t_0)] \\ Y_c = R_\oplus \sin\Theta \cdot \sin[\omega_z(t - t_0)] \\ Z_c = R_\oplus \cos\Theta \end{vmatrix}$

In (5) $\omega_\oplus$ is the Earth angular rotation velocity and $t_0$ is appropriately selected epoch (for example, the moment when the artificial earth satellite passes through the orbit perigee). If we substitute (5) in (4') we'll obtain:

(6) $A\cos\varphi + B\sin\varphi + C = 0$,

where $A = n_x \cdot \sin\Theta$, $B = n_y \cdot \sin\Theta$, $C = \sin\psi - n_z \cdot \cos\Theta$, $\varphi = \omega_\oplus(t - t_0)$.

By solving (6) we determine $\vec{R}_c$ at the tangenting moments $t_1$ and $t_2$ as well as the very moments. Thus, for the tangent vector we can write down:

(7) $\vec{\tau} = (\vec{R}_c \times \vec{n}) \times \vec{n}$

Vector $\vec{\tau}$ is determined in (7) in GeCS. We make a transformation of $\vec{\tau}$ in OCS [3]:

(8) $\vec{\tau}_{(OKS)} = \alpha_{OGe} \cdot \vec{\tau}_{(GeKS)}$

In (8) the transformation matrix $\alpha_{OGe}$ has the following form [3]:

$\alpha_{11} = \cos\omega \cdot \cos\Omega - \sin\omega \cdot \cos i \cdot \cos\Omega$
$\alpha_{12} = \cos\omega \cdot \sin\Omega + \sin\omega \cdot \cos i \cdot \cos\Omega$
$\alpha_{13} = \sin\omega \cdot \sin i$

$\alpha_{21} = -\sin\omega \cdot \cos\Omega - \sin\omega \cdot \cos i \cdot \sin\Omega$
$\alpha_{22} = -\sin\omega \cdot \sin\Omega + \sin\omega \cdot \cos i \cdot \cos\Omega$
$\alpha_{23} = \cos\omega \cdot \sin i$

$\alpha_{31} = \sin\Omega \cdot \sin i$
$\alpha_{32} = -\cos\Omega \cdot \sin i$
$\alpha_{33} = \cos i$

After determination of the tangent vector $\vec{\tau}$ in K, we can determine its crossing points with Keppler's ellipse in OCS:

(9) $\dfrac{(\xi+c)^2}{a^2} + \dfrac{\eta^2}{a^2(1-e^2)} = 1$, $\eta = k \cdot \xi$

In the second equation of system (9) k signifies the tangent's coefficient in OCS. The following relation exists between the orbital co-ordinates $(\xi,\eta)$ and the eccentric anomaly E [1]:

(10) $\begin{vmatrix} \xi = a(\cos E - e) \\ \eta = a\sqrt{1-e^2} \cdot \sin E \end{vmatrix}$,

where a is the large orbital semi-axis, e - is the eccentricity. On the other side, on the basis of Keppler's equation we can write down:

(11) $t = t_0 + (E - e \cdot \sin E)/\lambda$

After we find out the eccentric anomaly E in (10) and substitute it in (11), we determine the moments when the satellite crosses the specified tangents.

### 4. Estimation of the Method.

The explained method is analytical and it is presented by final formulae. It is reduced to a single application of the respective calculation procedure within the limits of one satellite circle. After correction of the orbital elements, the procedure can be repeated for the next interval of time.

The examined method is based on a situational condition whose geometrical model is reduced to the determination of tangents $\vec{\tau}_1$ and $\vec{\tau}_2$ in GeCS. The transformation of the tangents in OCS is equivalent to the transformation of the situational condition in the orbital plane [6].

A structural approach is applied for the method algorithmization. Based on a programme complex for situational analysis, developed for solution of the problems in [6], it was necessary to add two new sub-programmes for ensuring the treated situational problem. This means that the development of algorithms for situational analysis, based on the transformation of the situational conditions to Keppler's plane is facilitated by the presence of common sub-problems. In our case and for these in [6] this is the crossing of a straight line with Keppler's ellipse.

The following cases are possible for one Earth rotation around its axis:
- with sufficient orbital inclination equation (6) has four roots which leads to determination of four tangents connected with two crossings of segment $S$ with Keppler's plane;
- with smaller orbital inclination equation (6) has two solutions which determine two tangents, corresponding to one crossing of segment $S$ with $K$;
- with small orbital inclination segment $S$ doesn't cross $K$.

The correction of the orbital elements of each satellite circle on the basis of the selected model of disturbances allows to apply the presented approach for situational analysis within continuous interval of time. The method is applicable in the cases when Keppler's approximation in the terms of the satellite's circle period is admissible with a view to the solved problem. For solving practical problems in many cases this is executed.

**R e f e r e n c e s**